\documentclass[prl,aps,twocolumn]{revtex4}
\usepackage{times,epsfig,amssymb,amsfonts,amsmath}
\newcommand{\ket}[1]{|#1\rangle}
\newcommand{\bra}[1]{\langle #1|}

\usepackage{graphicx}
\usepackage{subfigure}
\usepackage{upgreek}
\usepackage[figuresright]{rotating}

\begin{document}

\title{Diagnosing Quantum Phases Using Long-Range Two-Site Quantum Resource Behaviors}

\author{Lin-Lin Su$^{1,2}$}
\author{Jun Ren$^{1,3}$}
\email{renjun@hebtu.edu.cn}
\author{Wen-Long Ma$^{2,4}$}
\email{wenlongma@semi.ac.cn}
\author{Z. D. Wang$^{3}$}
\email{zwang@hku.hk}
\author{Yan-Kui Bai$^{1,3}$}
\email{ykbai@semi.ac.cn}
\affiliation{$^1$ College of Physics and Hebei Key Laboratory of Photophysics Research and Application, Hebei Normal University, Shijiazhuang, Hebei 050024, China\\
$^2$ State Key Laboratory of Superlattices and Microstructures, Institute of Semiconductors, Chinese Academy of Sciences, Beijing 100083, China\\
$^3$ Department of Physics, HK Institute of Quantum Science \& Technology, and Guangdong-Hong Kong Joint Laboratory of Quantum Matter, The University of Hong Kong, Pokfulam Road, Hong Kong, China\\
$^4$ Center of Materials Science and Opto-Electronic Technology, University of Chinese Academy of Sciences, Beijing 100049, China}

\begin{abstract}
We propose and demonstrate that the behaviors of long-range, two-site quantum resources can effectively diagnose quantum phases. In an XX spin chain with symmetry-breaking quantum phase transitions, we reveal that the asymptotic and oscillating decay modes of quantum coherence or quantum discord, along with two-site distance, can identify two spin-liquid phases. Furthermore, based on our analytical results of spin correlation functions, we confirm the existence of long-range entanglement in the system and establish a connection between two-site entanglement and quantum phases. Additionally, for the extended Ising model with topological phase transitions, we find that coherence and quantum discord behaviors can also signify topological quantum phases. In particular, we discover the quantum resource freezing phenomenon, where topologically protected long-range quantum resources may have potential applications in quantum information processing.
\end{abstract}

\maketitle

\emph{Introduction.}---Quantum resources \cite{chi19rmp}, including quantum coherence, entanglement, and quantum discord, are fundamental to numerous quantum information processing (QIP) tasks \cite{rh09rmp, km12rmp, str17rmp, hu18physr}. Simultaneously, they have been developed into a powerful tool for studying physical properties in quantum many-body systems, with the relationship between quantum resources and quantum phases being a particularly fascinating problem \cite{ved08rmp, eis10rmp, chi18rpp, aban19rmp}.

Quantum phase transitions (QPT) in many-body systems are accompanied by dramatic changes in their physical properties \cite{sach11cam}, often exhibiting distinguishable signatures on quantum resources. In a piece of pioneering work, Osterloh \emph{et al} revealed the scaling behavior of two-site entanglement \cite{woo98prl} near the QPT point in the transverse Ising model \cite{ao02nat}. Beyond short-range, two-site entanglement, other long-range quantum resources such as quantum coherence \cite{ple14prl, abe06arxiv, yao15pra, nap16prl, fan16pra, kbu17prl}, quantum discord \cite{oll01prl, hen01jpa}, and multipartite entanglement \cite{cof00pra, mey02jmp, wei03pra, ou07pra, bai09pra, bai14prl, elt14jpa, bai21njp} can also characterize various spin models exhibiting symmetry-breaking QPTs \cite{tcw05pra, tro06pra, tro06prl, ycl11pra, sc13pra, mh14prb, gia14njp, spl18pra, mlh20pra, rrs21pra, lls22pra}, as well as topological quantum phase transitions (TQPTs) described by symmetry-protected topological order \cite{zeng19book, nr00prb, kit06ap, gev03, xgw04, xyf07prl, tre07prl, dhl07prl, hl08prl, gz15prl, wen17rmp, ver18prl, gong18prx}.

Although characterizing QPTs based on quantum resources has achieved significant success, identifying different quantum phases in many-body systems remains challenging. Recently, it was demonstrated that multipartite entanglement, evidenced by the scaling behavior of quantum Fisher information \cite{bra94prl, gio06prl, hyl12pra, tot12pra}, can characterize symmetry-protected topological phases in an extended Kitaev chain \cite{lp17prl, yrz18prl}. Furthermore, through varying robustness of von Neumann entanglement against parameter perturbations, topological phases can be experimentally determined in split-step quantum walks \cite{qqw20op}. Compared to multipartite entanglement, two-site quantum entanglement, quantum coherence, and quantum correlation are fundamental quantum resources in QIP that are more easily detected experimentally. However, it remains unexplored whether the behaviors of such long-range two-site quantum resources can identify different quantum phases in many-body systems.

In this Letter, we examine the behaviors of long-range two-site quantum resources and demonstrate that the decay modes of these resources, along with the two-site distance, can serve as diagnostics for quantum phases. In an XX chain with three-spin interactions \cite{ycl11pra} exhibiting symmetry-breaking QPTs, we show that the asymptotic and oscillating decay modes of two-site quantum coherence and quantum correlations \cite{oll01prl, hen01jpa} can identify two types of spin-liquid phases. The behavior of long-range two-site entanglement \cite{woo98prl} is also confirmed and exhibits the same functionality. Additionally, in the extended Ising model with TQPTs \cite{gz15prl}, we explore the relationship between two-site quantum resources and topological quantum phases by uncovering a quantum resource freezing phenomenon, which can characterize the topological quantum phases with winding numbers $\mathcal{N}=\pm 1$.

\emph{The extended Ising model and its two-site quantum states.}---We study the extended Ising model with three-spin interactions \cite{gz15prl},
\begin{eqnarray}
H=&-&\underset{j=1}{\overset{L}{\sum}}[\frac{1+\gamma}{2}\sigma^{x}_{j}\sigma^{x}_{j+1}+\frac{1-\gamma}{2}\sigma^{y}_{j}\sigma^{y}_{j+1}+\lambda\sigma^{z}_{j}\nonumber\\
&+&\alpha\sigma^{z}_{j}(\frac{1+\delta}{2}\sigma^{x}_{j-1}\sigma^{x}_{j+1}+\frac{1-\delta}{2}\sigma^{y}_{j-1}\sigma^{y}_{j+1})],
\end{eqnarray}
where $L$ is total spin number in the spin chain, $\gamma$ is the anisotropic parameter of the nearest-neighbor couplings, $\lambda$ is the strength of external magnetic field, $\alpha$ and $\delta$ represent the strength and anisotropy of three-spin interactions, and $\{\sigma^{x}_{j}$, $\sigma^{y}_{j}$ and $\sigma^{z}_{j}\}$ are the Pauli operators on the $j$th spin. When $\gamma=\delta=0$, the system is the XXT model (an XX chain with isotropic three-spin interactions) exhibiting the symmetry-breaking QPTs. In a generic case for $\gamma\neq 0$ and $\delta\neq 0$, the extended Ising model exhibits the topological QPTs.

The Hamiltonian in Eq. (1) can be exactly diagonalized and a review on the details is presented in the Supplemental Material \cite{sup23bai}. For the $i$th and $j$th spins, the two-site reduced density matrix of the ground state has the form,
\begin{equation}
	\begin{split}
&\rho_{r}=\rho_{ij}=\begin{pmatrix} u^{+} & 0 & 0 & y^{-}  \\
0 & z & y^{+} & 0 \\
0 & y^{+} & z & 0 \\
y^{-} & 0 & 0 & u^{-} \end{pmatrix},
	\end{split}
\end{equation}
where $r=|j-i|$ denotes the distance between the two spins, and the nonzero matrix elements are  $u^{\pm}=(1\pm 2\langle\sigma^z\rangle+\langle\sigma^z_0\sigma^z_r\rangle)/4$, $z=(1-\langle\sigma^z_0\sigma^z_r\rangle)/4$ and $y^{\pm}=(\langle\sigma^x_0\sigma^x_r\rangle\pm\langle\sigma^y_0\sigma^y_r\rangle)/4$. The magnetization $\langle\sigma^z\rangle$ and two-site correlations $\langle\sigma^s_0\sigma^s_r\rangle$ with $s=x,y,z$ are the functions of $G_r$ \cite{sup23bai}. For a finite chain length $L$ at zero temperature, $G_r$ function can be written as \cite{be70pra,be71pra}
\begin{equation}
 G_{r}=-\frac{1}{L}{\underset{k}{\sum}}\frac{1}{\Lambda_{k}}\{\cos(\phi_k r)z_k+\sin(\phi_k r)y_k\},
\end{equation}
where $\phi_k=2\pi k/L$ with $k=-M, \dotsc, M$ and $M=(L-1)/2$ for odd $L$, the energy spectra $\Lambda_k=\sqrt{z_k^2+y_k^2}$ with $z_k=\lambda-\cos\phi_k-\alpha\cos(2\phi_k)$ and $y_k=\gamma\sin\phi_k+\alpha\delta\sin(2\phi_k)$, respectively. In the thermodynamic limit, the function has the following integral form \cite{el61ap}
\begin{equation}
	\begin{split}
		G_{r}=-&\frac{1}{\pi}\int^{\pi}_0 F(\gamma,\lambda,\alpha, \delta, r)d\phi,
	\end{split}
\end{equation}
where the kernel is $F(\gamma,\lambda,\alpha, \delta, r)=[\cos(\phi r)z+\sin(\phi r)y]/\Lambda_{\phi}$ with the parameters being $\Lambda_{\phi}=\sqrt{z^2+y^2}$, $z=\lambda-\cos\phi-\alpha\cos(2\phi)$, and $y=\gamma\sin\phi+\alpha\delta\sin(2\phi)$. The $G_r$ function here is vital in calculating quantum state $\rho_r$ and analyzing the behaviors of two-site quantum resources.

\emph{Characterization of quantum phases in the XXT model.}---We first consider the XXT model for which the Hamiltonian in Eq. (1) has the anisotropic parameters $\gamma=0$ and $\delta=0$ \cite{ycl11pra}. In this case, the ground state phase diagram is composed of ferromagnetic phase, spin-liquid I (SL-I) phase and spin-liquid II (SL-II) phase \cite{it03epjb}. The symmetry-breaking QPTs are connected with the number of Fermi points \cite{it03epjb,mf96prb,ra98prb,sd00prb,ca01prl,mo15prb}. By solving the energy equation $\Lambda_{\phi}=0$ with $\phi\in \{0,\pi\}$ in the thermodynamic limit, we obtain the two real Fermi points
\begin{equation}
\phi_{\pm}=\arccos\left[\frac{-1\pm\sqrt{1+8\alpha^2+8\alpha\lambda}}{4\alpha}\right].
\end{equation}
As shown in Fig. 1, the Fermi points are plotted as the functions of magnetic field $\lambda$ and interaction strength $\alpha$, where the cyan surface is $\phi_-$, the orange surface is $\phi_+$, the green plane is the background. We find the number of Fermi points can not only detect the QPTs but also identify the three quantum phases, where there are two Fermi points in the SL-II phase, one in the SL-I phase and zero in ferromagnetic phase.

\begin{figure}
	\epsfig{figure=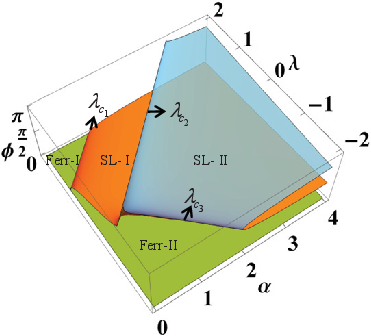,width=0.42\textwidth}
\caption{(Color online) The ground state phase diagram of the XXT model described by the number of Fermi points, where the cyan surface is $\phi_-(\lambda,\alpha)$, the orange surface is $\phi_+(\lambda,\alpha)$, the green plane is the background. The critical lines are $\lambda_{c_1}=\alpha+1$, $\lambda_{c_2}=\alpha-1$ and $\lambda_{c_3}=-(1+8\alpha^2)/8\alpha$ ($\alpha\geq 0.25$ for $\lambda_{c_3}$), and the regions of Ferr-I and Ferr-II are the same phase.}
\end{figure}

We now study the property of long-range two-site quantum coherence of reduced state $\rho_r$ along with the distance of two spins in the XXT spin chain. Quantum coherence is an important resource in the QIP \cite{mh16pra,hl17pra,cn16prl} and quantum thermodynamics \cite{mh13nc,fg13prl,gg15pr,yang22prl}. Here we adopt the $l_1$-norm quantum coherence \cite{ple14prl}, which is expressed as the sum of modulus of off-diagonal element: $C_{l_1}(\rho_r)=\sum_{i,j,i\neq j}|\rho_{ij}|$. In the XXT model, since the Hamiltonian is isotropic in $x$ and $y$ direction ($\gamma=\delta=0$), we have the matrix element $y^-=(\langle\sigma^x_0\sigma^x_r\rangle-\langle\sigma^y_0\sigma^y_r\rangle)/4=0$ in Eq. (2). Therefore, the $l_1$-norm coherence of two-site quantum state is
\begin{equation}
	 C_{l_1}(\rho_r)=2|y^{+}|=|\langle\sigma^x_0\sigma^x_r\rangle|,
\end{equation}
where the spin correlation $\langle\sigma^x_0\sigma^x_r\rangle$ is a determinant composed of $G_r$ function \cite{sup23bai}. In order to obtain the quantum coherence expediently, we derive the analytical formula of $G_r$ function in the thermodynamic limit for different regions of quantum phases (see the Supplemental Material \cite{sup23bai})
\begin{equation}
	\begin{split}
		G_r=\begin{cases}
			-\delta_{0r},& \rm{Ferr-I}\\
			\delta_{0r},& \rm{Ferr-II}\\
			\frac{2\sin (r\phi_{+})}{\pi r}-\delta_{0r},& \rm{SL-I}\\
			2(\frac{\sin (r\phi_{+})}{\pi r}-\frac{\sin (r\phi_{-})}{\pi r})+\delta_{0r},& \rm{SL-II}
		\end{cases}
	\end{split}
\end{equation}
where $r$ is the distance of two spins and  $\phi_{\pm}$ are the Fermi points given in Eq. (5). According to this analytical formula, the spin correlation $\langle\sigma^x_0\sigma^x_r\rangle$ is zero for the regions of Ferr-I and Ferr-II in Fig. 1, so the quantum coherence $C_{l_1}(\rho_r)$ is zero in the ferromagnetic phase. However, the quantum coherence in the SL-I and SL-II phases is nonzero in general. As shown in Fig. 2(a)-(c), we plot the coherence $C_{l_1}(\rho_r)$ as a function of parameters $\lambda$ and $\alpha$ for spin-spin distances $r=1,6$ and $10$. We can distinguish the SL-I and SL-II phases from the ferromagnetic phase, and the change patterns of $C_{l_1}$ along with the parameters $\alpha$ and $\lambda$ are different as shown in panels (a)-(c), where $C_{l_1}$ is asymptotical in SL-I phase and oscillating in SL-II phase. The intrinsic reason is that the coherence decays along with the distance $r$ in the asymptotical mode for SL-I phase but in the oscillating mode for SL-II phase as shown in Fig. 2(d).

\begin{figure}
	\epsfig{figure=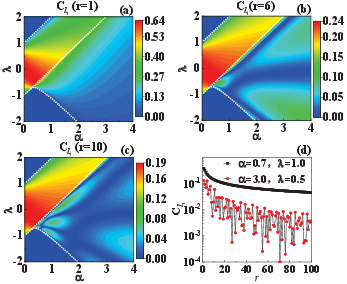,width=0.48\textwidth}
\caption{(Color online) The $l_1$-norm coherence $C_{l_1}(\alpha, \lambda)$ for different two-site distances: (a) $r=1$, (b) $r=6$, and (c) $r=10$, respectively. In panel (d), the parameters $(\alpha=0.7,\lambda=1.0)$ in SL-I phase and $(\alpha=3.0,\lambda=0.5)$ in SL-II phase are chosen, where the asymptotical and oscillating decay modes of $C_{l_1}$ can serve as the diagnostic of quantum phases.}
\end{figure}

Then we study the long-range two-site quantum entanglement in the same model. As quantified by concurrence \cite{woo98prl}, the two-site entanglement can be written as
\begin{equation}
C(\rho_r)={\rm{max}}\{0,2(|y^{+}|-\sqrt{u^{+}u^{-}})\},
\end{equation}
where $y^+$ and $u^{\pm}$ are the off-diagonal and diagonal elements of reduced quantum state $\rho_r$ in Eq. (2), respectively. In comparison with the $l_1$-norm coherence in Eq. (6), the two-site entanglement is less than the corresponding quantum coherence. Moreover, although quantum coherence has the long-range property, two-site entanglement is often short-range and experiences entanglement sudden death (ESD) \cite{kz01pra,ss03jmo,ty04prl,mpa07sci,jl04prl,ty09sci} in multipartite spin systems without long-range interactions \cite{ao02nat,lls22pra}. We analyze the property of two-site concurrence in the XXT model, and confirm the existence of long-range two-site entanglement, for which the parameters are close to the critical lines $\lambda_{c_1}$, $\lambda_{c_2}(\alpha\leq 0.25)$, and $\lambda_{c_3}$ \cite{sup23bai}. In SL-I and SL-II phases, we choose three sets of parameters near the three critical lines, and calculate the concurrence as a function of two-site distance. As shown in Fig. 3, the decay modes of long-range two-site entanglement have the same functionality for identifying quantum phases, where $C_{l_1}$ decays in the asymptotical mode for SL-I phase but in the oscillating mode for SL-II phase.

\begin{figure}
	\epsfig{figure=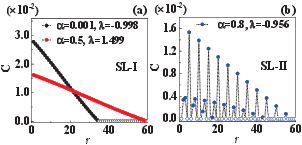,width=0.48\textwidth}
	\caption{(Color online) Long-range two-site entanglement in (a) SL-I phase and (b) SL-II phase, where the distinctly different decay modes of concurrence serve as the diagnostic of quantum phases. The hollow dots indicate the distances with zero entanglement.}
\end{figure}

\emph{Characterization of topological quantum phases in the extended Ising model.}--- In the extended Ising system with $\gamma\neq 0$ and $\delta\neq 0$, there are topological quantum phases in the ground state, for which the winding number is an effective geometric order parameters \cite{gz15prl}. The Hamiltonian in Eq. (1) can be further expressed as the Bogoliubov-de Gennes form in the momentum space \cite{gz15prl,yrz18prl}
\begin{equation}
H=\underset{k=-M}{\overset{M}{\sum}}\begin{pmatrix}c^{\dagger}_{k}&c_{-k}\end{pmatrix}\mathcal{H}_{k}\begin{pmatrix}c_{k}\\c^{\dagger}_{-k}\end{pmatrix},
\end{equation}
where $M=(L-1)/2$ for odd $L$, $\mathcal{H}_{k}=\vec{r}(k)\cdot \vec{\sigma}$ with $\vec{r}(k)=[0,y_k,z_k]$ and $\vec{\sigma}=(\sigma^x,\sigma^y,\sigma^z)$, and the vector $\vec{r}$ represents a two-dimensional magnetic field with the components $y_k=\gamma\sin\phi_k+\alpha\delta\sin(2\phi_k)$ and $z_k=\lambda-\cos\phi_k-\alpha\cos(2\phi_k)$. Zhang and Song established the connection between topological quantum phase and corresponding winding number \cite{gz15prl}
\begin{eqnarray}
\mathcal{N}=\frac{1}{2\pi}\oint(ydz-zdy)/|\vec{r}|^2,
\end{eqnarray}
which is an integer representing the direction and total times that the closed loop travels around the origin point in the $y$-$z$ plane \cite{rbs02GT}. The winding numbers can identify topological quantum phases in the extended Ising model, and the critical points for the TQPTs are available by solving the characteristic equations \cite{spl18pra}.

\begin{figure}
	\epsfig{figure=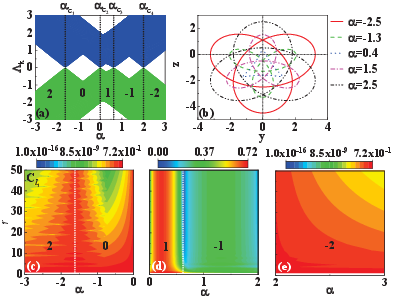,width=0.48\textwidth}
\caption{(Color online) Energy spectra, winding vector trajectories, and long-range quantum coherence $C_{l_1}(\alpha, r)$ in the extended Ising model with the parameters $\gamma=1$, $\delta=-1$, and $\lambda=1$ for the chain length $L=1001$. (a) The energy spectra as a function of three-spin interaction $\alpha$ where the four critical points and five topological phases are labeled. (b) The trajectories of the winding vectors in the $y$-$z$ plane which correspond to the winding numbers $2$, $0$, $1$, $-1$ and $-2$, respectively. (c)-(e) The behaviors of long-range two-site quantum coherence in different topological quantum phases.}
\end{figure}

Next, we study the connection between the long-range two-site quantum coherence and topological quantum phases in the extended Ising model. When the TQPTs are driven by the three-spin interaction $\alpha$, we choose the other parameters in the Hamiltonian to be $\gamma=1$, $\delta=-1$, and $\lambda=1$, respectively. As shown in Fig. 4(a), the energy spectra $\Lambda_k=\pm\sqrt{z_k^2+y_k^2}$ are plotted for a finite chain length $L=1001$, where the critical points $\alpha_{c_1}=(-\sqrt{5}-1)/2$, $\alpha_{c_2}=0$, $\alpha_{c_3}=(\sqrt{5}-1)/2$, and $\alpha_{c_4}=2$ are labeled. The trajectories of the typical winding vectors in the auxiliary $y$-$z$ plane are plotted in Fig.~4(b). As quantified by the $l_1$-norm measure, the two-site quantum coherence in the extended Ising system is
\begin{equation}
C_{l_1}(\rho_r)=2(|y^+|+|y^-|),
\end{equation}
where $y^\pm=(\langle\sigma^x_0\sigma^x_r\rangle\pm\langle\sigma^y_0\sigma^y_r\rangle)/4$ are the off-diagonal elements of two-site quantum state $\rho_r$ in Eq. (2). It is noted that the first-order derivative of adjacent two-site coherence $C_{l_1}$ can detect all the TQPTs in the extended Ising system \cite{sup23bai}. Here we focus on the properties of the long-range two-site coherence in different topological phases. In Fig. 4(c)-(e), we plot $C_{l_1}(\rho_r)$ as a function of three-spin interaction strength $\alpha$ and two-site distance $r$, where the different topological phases are labeled by the corresponding winding numbers. In the case of $\mathcal{N}=2$, the coherence decays in the oscillating mode along with the increase of two-site distance for a given value of $\alpha$. But, in the case of $\mathcal{N}=-2$, the coherence decays in the asymptotical mode along with two-site distance, although there exist initial oscillating for short two-site distance due to the influence of adjacent three-spin interactions. Moreover, the strength of $|\alpha|$ can change the decay rate in both topological phases of $\mathcal{N}=\pm 2$. In the case of $\mathcal{N}=0$, the coherence can decay in both oscillating and asymptotical modes for different values of $\alpha$ as shown in Fig. 4(c). We note that the ground state of the XXT model we analyzed previously is also a topological trivial ($\mathcal{N}=0$) where the quantum coherence exhibits two kinds of decay modes too.

In the topological phases with winding numbers $\mathcal{N}=\pm 1$, the behavior of two-site quantum coherence exhibits a remarkable property, where the nonzero coherence will become stable for a given value of $\alpha$ after a slight decay of short two-site distances as shown in Fig. 4(d). In addition, the higher steady coherence in the phase $\mathcal{N}=1$ comes from spin correlation $\langle \sigma^y_0\sigma^y_r\rangle$, while the lower steady coherence in the phase $\mathcal{N}=-1$ is due to $\langle \sigma^x_0\sigma^x_r\rangle$ (see the Supplemental Material \cite{sup23bai}). The steady quantum coherence is a kind of topologically protected long-range quantum resource, and we refer to it as quantum coherence freezing (QCF) phenomenon along with the two-site distance, which is different from the case of frozen quantum coherence under dynamical conditions \cite{bro15prl,tong16pra,fan16sr,sil16prl}. Therefore, as shown in Fig. 4, the behaviors of long-range two-site coherence, such as oscillating and asymptotical decay modes or the QCF mode, can serve as an effective diagnostic of topological quantum phases. In addition, we also investigate the topological quantum phases driven by the external magnetic field $\lambda$, and establish the connection between long-range quantum coherence and the topological phases, where the coherence still exhibits the QCF phenomenon in the case $\mathcal{N}=1$ \cite{sup23bai}.

The topologically protected steady coherence is a kind of useful quantum resource in the QIP. In order to explain further the QCF phenomenon in the extended Ising model, we consider a special case of the Hamiltonian with $\gamma=1$ and $\lambda=\alpha=0$, for which the winding number is $\mathcal{N}=1$ and its ground state has an analytical form \cite{gz15prl}
\begin{equation}
\ket{G_1}=\frac{1}{\sqrt{2}}(\underset{j\in e}{\Pi}\ket{\nearrow}_j\underset{j\in o}{\Pi}\ket{\swarrow}_j+\underset{j\in e}{\Pi}\ket{\swarrow}_j\underset{j\in o}{\Pi}\ket{\nearrow}_j),	
\end{equation}
where $\sigma^x_{j}\ket{\nearrow}_j(\ket{\swarrow}_j)=\ket{\nearrow}_j(-\ket{\swarrow}_j)$, and the summing targets $e$ and $o$ denote the even and odd number of sites, respectively. Its reduced quantum state of the $i$th and $j$th spins with distance $r$ is
\begin{equation}
	\begin{split}
		&\rho_{r}^{\mathcal{N}=1}=\frac{1}{4}\begin{pmatrix} 1 & 0 & 0 & -1  \\
0 & 1 & -1 & 0 \\
0 & -1 & 1 & 0 \\
-1 & 0 & 0 & 1 \end{pmatrix},
	\end{split}
\end{equation}
which has the steady coherence $C_{l_1}(\rho_{r}^{\mathcal{N}=1})=1$ and is a constant matrix independent of the two-site distance. Similarly, for the topological phase $\mathcal{N}=-1$ with the parameters $\gamma=-1$ and $\lambda=\alpha=0$, we derive the distance-independent two-site quantum state $\rho_{r}^{\mathcal{N}=-1}$ via the analytical ground state $\ket{G_{-1}}$ \cite{gz15prl}, which exhibits the QCF phenomenon with topologically protected quantum coherence $C_{l_1}(\rho_{r}^{\mathcal{N}=-1})=1$.

\emph{Discussions.}--- The above results for quantum coherence and quantum entanglement can be extended to other two-site quantum resource, such as quantum discord that can describe quantum correlations beyond quantum entanglement \cite{mod10prl,str12prl,bai13pra}. The two-qubit quantum discord can be quantified by the formula \cite{oll01prl,hen01jpa}
\begin{equation}
 D_{A}(\rho_{AB})=S(\rho_{A})-S(\rho_{AB})+\underset{\{E^A_{k}\}}{\rm{min}}\underset{k}{\sum}p_{k}S(\rho_{B|k}),
\end{equation}
where the measurement $\{E_k^A\}$ is performed on subsystem $A$ with the minimum running over all the projection measurements, and $S(\cdot)$ is the von Neumann entropy. Quantum discord can characterize quantum advantage without entanglement and is a kind of key resource in the QIP \cite{kni98prl,lan08prl,dat08prl,roa11prl,dak12natp}. In the XXT model, the behaviors of long-range two-site quantum discord $D_{A}(\rho_r)$ is further studied, where we find that the discord $D_{A}(\rho_{r})$ decays along with two-site distance in the asymptotical mode for the SL-I phase and in the oscillating mode for the SL-II phase \cite{sup23bai}, which indicates that quantum discord has the same functionality for identifying quantum phases as that of quantum coherence. Furthermore, in the extended Ising model, we find two-site entanglement is short-range but quantum discord is long-range which exhibits behaviors similar to those of quantum coherence \cite{sup23bai}. In particular, unlike the frozen quantum discord in dynamical procedure \cite{maz10prl,you12pra}, we find a novel phenomenon of quantum discord freezing in topological phases with the winding numbers $\mathcal{N}=\pm 1$.

\emph{Conclusions and outlook.}---In summary, we have investigated the behaviors of long-range two-site quantum resources and uncovered the connection between their decay modes and various quantum phases. In the XXT model \cite{ycl11pra} with symmetry-breaking QPTs, we demonstrate that the asymptotic or oscillating decay modes of two-site quantum coherence and quantum discord can identify two spin-liquid phases. We also confirm the existence of long-range two-site entanglement in this model, which can likewise signify different quantum phases. In the extended Ising model \cite{gz15prl}, we show that the decay modes of long-range two-site quantum resources can serve as an effective diagnostic tool for topological phases. Compared to Fisher information or multipartite entanglement of the global ground state in many-body systems, the long-range resource behaviors of a set of two-site quantum states are more easily detected experimentally. Lastly, the topologically protected two-site quantum resources exhibit freezing phenomena, which hold promising applications in both quantum communication and quantum metrology \cite{pez18rmp, por22rmp}.

This work was supported by NSF-China (Grants Nos. 11575051 and 11904078), Hebei NSF (Grants Nos. A2021205020, A2019205266), and project of China Postdoctoral Science Foundation (Grant No. 2020M670683). W.L.M. acknowledges support from the Chinese Academy of Sciences (No. E0SEBB11, No. E27RBB11) and the National Natural Science Foundation of China (No. 121743379, No. E31Q02BG).

\newpage

\setcounter{equation}{0}
\setcounter{figure}{0}

\section{Supplemental material}

\subsection{I. The diagonalization procedure for the Hamiltonian of the extended Ising model}

The Hamiltonian of the extended Ising model given in the main text has the form \cite{gz2015prl}
\begin{eqnarray}
H=&-&\underset{j=1}{\overset{L}{\sum}}[\frac{1+\gamma}{2}\sigma^{x}_{j}\sigma^{x}_{j+1}+\frac{1-\gamma}{2}\sigma^{y}_{j}\sigma^{y}_{j+1}+\lambda\sigma^{z}_{j}\nonumber\\
	 &+&\alpha\sigma^{z}_{j}(\frac{1+\delta}{2}\sigma^{x}_{j-1}\sigma^{x}_{j+1}+\frac{1-\delta}{2}\sigma^{y}_{j-1}\sigma^{y}_{j+1})],
\end{eqnarray}
where $L$ is the total spin number in the spin chain, $\gamma$ is the anisotropic parameter of the nearest-neighbor couplings, $\lambda$ is the strength of external magnetic field, $\alpha$ and $\delta$ represent the strength and anisotropy of three-spin interactions, and $\sigma^{x}_{j}$, $\sigma^{y}_{j}$ and $\sigma^{z}_{j}$ are the Pauli operators on the $j$th spin. Next, we will give a brief review on the diagonalization procedure for this Hamiltonian \cite{el1961ap,pp70ap,eb1971pra}.

The spin operators in the Hamiltonian can be mapped into the spinless fermion operators by introducing the Jordan-Wigner transformation \cite{sac2011cam}
\begin{eqnarray}
	&&\sigma^{z}_{j}=1-2c^{\dagger}_{j}c_{j},\nonumber\\
	&&\sigma^{x}_{j}={\underset{l<j}{\prod}}(1-2c^{\dagger}_{l}c_{l})(c^{\dagger}_{j}+c_{j}),\nonumber\\
	&&\sigma^{y}_{j}=-i{\underset{l<j}{\prod}}(1-2c^{\dagger}_{l}c_{l})(c^{\dagger}_{j}-c_{j}),
\end{eqnarray}
where $c^{\dagger}_{j}$ and $c_{j}$ are the spinless creation and annihilation operators. After substituting the expressions in Eq. (2) into the Hamiltonian in Eq. (1), we can obtain the following form
\begin{eqnarray}
	H=&-&\underset{j=1}{\overset{L}{\sum}}\lambda(1-2c^{\dagger}_{j}c_{j})\nonumber\\
      &-&\underset{j=1}{\overset{L}{\sum}}[\gamma(c^{\dagger}_{j}c^{\dagger}_{j+1}-c_{j}c_{j+1})+c^{\dagger}_{j}c_{j+1}-c_{j}c^{\dagger}_{j+1}]\nonumber\\
	  &-&\underset{j=1}{\overset{L}{\sum}}\alpha(c^{\dagger}_{j-1}c_{j+1}-c_{j-1}c^{\dagger}_{j+1})\nonumber\\
	  &-&\underset{j=1}{\overset{L}{\sum}}\alpha\delta(c^{\dagger}_{j-1}c^{\dagger}_{j+1}+c_{j-1}c_{j+1}).
\end{eqnarray}
Furthermore, the above Hamiltonian can be expressed in the momentum space by the Fourier transformation, and the Fourier transformation of fermion annihilation operator can be written as
\begin{equation}
c_{j}=\frac{1}{\sqrt{L}}\underset{k=-M}{\overset{M}{\sum}}c_{k}\exp(-ij\phi_{k}),
\end{equation}
where $\phi_k=2\pi k/L$ with $k=-M,\dotsc,M$ and $M=(L-1)/2$ for odd $L$. After some derivation, we can rewrite the Hamiltonian in the following form
\begin{equation}
	\begin{split}
		&H=-\lambda L+2(\lambda-1-\alpha)c^{\dagger}_{0}c_{0}\\
		 &+\underset{k=1}{\overset{M}{\sum}}2[\lambda-\cos\phi_{k}-\alpha\cos(2\phi_{k})](c^{\dagger}_{k}c_{k}+c^{\dagger}_{-k}c_{-k})\\
        &+\underset{k=1}{\overset{M}{\sum}}2i[\gamma\sin\phi_{k}+\alpha\delta\sin(2\phi_{k})](c^{\dagger}_{k}c^{\dagger}_{-k}+c_{k}c_{-k}).
	\end{split}
\end{equation}

After further applying the Bogoliubov transformation, the Hamiltonian in Eq. (5) can be transformed into the diagonal form
\begin{eqnarray}
	H=\underset{k=-M}{\overset{M}{\sum}}2\Lambda_{k}(\eta^{\dagger}_{k}\eta_{k}-\frac{1}{2}),
\end{eqnarray}
for which the energy spectra are $\pm\Lambda_{k}$ and
\begin{eqnarray}
	\Lambda_{k}=\sqrt{z(k)^2+y(k)^2}
\end{eqnarray}
with $z(k)=\lambda-\cos\phi_{k}-\alpha\cos(2\phi_{k})$ and $y(k)=\gamma\sin\phi_{k}+\alpha\delta\sin(2\phi_{k})$, and the mapping operator of Bogoliubov transformation is
\begin{eqnarray}
	\eta_{k}=\cos\frac{\theta_{k}}{2}c_{-k}-i\sin\frac{\theta_{k}}{2}c^{\dagger}_{k},
\end{eqnarray}
in which the coefficients are
\begin{eqnarray}
	\sin\frac{\theta_{k}}{2}&=&\frac{\Lambda_{k}-z(k)}{\sqrt2\Lambda_{k}[\Lambda_{k}-z(k)]},\nonumber\\
	\cos\frac{\theta_{k}}{2}&=&\frac{y(k)}{\sqrt2\Lambda_{k}[\Lambda_{k}-z(k)]},
\end{eqnarray}
with the parameter being
\begin{eqnarray}
	\theta_{k}&=&\arcsin\{-[\gamma\sin\phi_{k}+\alpha\delta\sin(2\phi_{k})]/\Lambda_{k}\}.
\end{eqnarray}

\subsection{II. Two-site reduced density matrix of ground state in the extended Ising model}

In the extended Ising model, the two-site reduced density matrix of ground state for the $i$th and $j$th spins given in the main text has the form
\begin{equation}
	\begin{split}
		&\rho_{r}=\rho_{ij}=\begin{pmatrix} u^{+} & 0 & 0 & y^{-}  \\
			0 & z & y^{+} & 0 \\
			0 & y^{+} & z & 0 \\
			y^{-} & 0 & 0 & u^{-} \end{pmatrix},
	\end{split}
\end{equation}
where $r=|j-i|$ denotes the distance between the two spins, and the nonzero matrix elements are $u^{\pm}=(1\pm 2\langle\sigma^z\rangle+\langle\sigma^z_0\sigma^z_r\rangle)/4$, $z=(1-\langle\sigma^z_0\sigma^z_r\rangle)/4$ and $y^{\pm}=(\langle\sigma^x_0\sigma^x_r\rangle\pm\langle\sigma^y_0\sigma^y_r\rangle)/4$, which can be calculated from the magnetization and spin correlation functions.

For the nonzero elements in $\rho_r$, the magnetization $\langle\sigma^z\rangle$ and two-site correlation functions $\langle\sigma^s_0\sigma^s_r\rangle$ with $s=x,y,z$ can be written as \cite{eb1971pra,eb1970pra}
\begin{eqnarray}
		&&\langle\sigma^z\rangle=G_0,\nonumber\\
		&&\langle\sigma^x_0\sigma^x_r\rangle=\begin{vmatrix} G_{-1} & G_{-2} & \cdots & G_{-r} \\ G_{0} & G_{-1} & \cdots & G_{-r+1} \\ \vdots & \vdots & \ddots  & \vdots \\  G_{r-2} & G_{r-3} & \cdots & G_{-1} \end{vmatrix},\nonumber\\
		&&\langle\sigma^y_0\sigma^y_r\rangle=\begin{vmatrix} G_{1} & G_{0} & \cdots & G_{-r+2} \\ G_{2} & G_{1} & \cdots & G_{-r+3} \\ \vdots & \vdots & \ddots  & \vdots \\ G_{r} & G_{r-1} & \cdots & G_{1} \end{vmatrix},\nonumber\\
		&&\langle\sigma^z_0\sigma^z_r\rangle=\langle\sigma^z\rangle^2-G_{r}G_{-r},
\end{eqnarray}
where $G_r$ function for the case of a finite chain length has the expression given in Eq. (3) of main text and its formula for the case of the thermodynamic limit is presented in Eq. (4) of main text. Therefore, according to Eqs. (11) and (12), the two-site quantum state $\rho_r$ can be obtained by calculating the $G_r$ functions and related determinants. It should be noted that, in order to analyze the properties of quantum state and its quantum resource expediently, an analytical formula for the $G_r$ function is very desirable.

\subsection{III. The analytical formula for the $G_r$ function in the XXT model}

We consider the XXT model, for which the Hamiltonian shown in Eq. (1) of the main text has the parameters $\gamma=0$, $\delta=0$, and  $\lambda \neq 0$. In the thermodynamic limit, the $G_r$ function has the form \cite{el1961ap}
\begin{equation}
	\begin{split}
		G_{r}=-&\frac{1}{\pi}\int^{\pi}_0 F(\lambda,\alpha, r)d\phi,
	\end{split}
\end{equation}
where the integral kernel is
\begin{equation}
F(\lambda,\alpha, r)=\cos(\phi r)z/\Lambda_{\phi}
\end{equation}
in which the parameter is $\Lambda_{\phi}=\sqrt{z^2}$ with $z=\lambda-\cos\phi-\alpha\cos(2\phi)$.

The Fermi points \cite{it2003epjb,mf1996prb} characterize the positions at which the energy $\Lambda_{\phi}$ is equal to zero, which plays a key role in the analysis on different quantum phases in the XXT model as shown in Fig. 1 of the main text. There are two Fermi points in the SL-II phase, one in the SL-I phase and zero in the Ferromagnetic phase (the regions of Ferr-I and Ferr-II are the same phase). By solving the energy equation $\Lambda_{\phi}=0$ with $\phi\in\{0,\pi\}$ in the thermodynamic limit, we can obtain the two Fermi points $\phi_{+}$ and $\phi_{-}$ which can be written as
\begin{equation}
	\begin{split}
		\phi_{\pm}=\arccos[\frac{-1\pm\sqrt{1+8\alpha^2+8\alpha\lambda}}{4\alpha}].
	\end{split}
\end{equation}
It should be noted that the Fermi points are required to be real numbers.

Next, according to the numbers and values of Fermi points, we will analyze the kernel functions $F(\lambda,\alpha, r)$ in different quantum phases. In the ferromagnetic phase, there is no Fermi point and we can demonstrate that $z(\phi)=\lambda-\cos\phi-\alpha\cos(2\phi)$ is positive in the region of Ferr-I and negative in the region of Ferr-II. Therefore, we have
\begin{eqnarray}
	F(\lambda,\alpha, r)&=&\frac{\cos (r\phi)[\lambda-\cos(\phi)-\alpha\cos(2\phi)]}{\sqrt{[\lambda-\cos(\phi)-\alpha\cos(2\phi)]^2}}\nonumber\\
	&=&\begin{cases}
		\cos (r\phi),& \rm{Ferr-I}\\
		-\cos (r\phi).& \rm{Ferr-II}
	\end{cases}
\end{eqnarray}
Substituting the result into the $G_r$ function in Eq. (13), we can derive that $G_r=0$ for the case of $r\neq 0$ and the nonzero $G_r$ corresponds to the case of $r=0$. After some calculation, we obtain the analytical formula
\begin{eqnarray}
	G_{r}&=&\begin{cases}
		-\delta_{0r},& \rm{Ferr-I}\\
		\delta_{0r}.& \rm{Ferr-II}
	\end{cases}
\end{eqnarray}

In the SL-I phase, there is only one Fermi point $\phi_+$ which divides the function $z(\phi)$ into two parts. It can be demonstrated that $z(\phi)$ is negative in the region $\phi\in [0,\phi_+)$ and positive in the region $\phi\in [\phi_+, \pi]$. Therefore, according to Eq. (14), we can have the kernel function
\begin{eqnarray}
	F(\lambda,\alpha, r)&=&\frac{\cos (r\phi)[\lambda-\cos(\phi)-\alpha\cos(2\phi)]}{\sqrt{[\lambda-\cos(\phi)-\alpha\cos(2\phi)]^2}}\nonumber\\
	&=&\begin{cases}
		-\cos (r\phi),& \phi\in[0,\phi_{+})\\
		\cos (r\phi).& \phi\in[\phi_{+},\pi]
	\end{cases}
\end{eqnarray}
By substituting the above expression into Eq. (13), we can derive the analytical formula of $G_r$. In the case of $r\neq 0$, we have
\begin{eqnarray}
	G_{r}&=&\frac{1}{\pi}\int^{\phi_{+}}_0 \cos (r\phi)d\phi-\frac{1}{\pi}\int^{\pi}_{\phi_{+}} \cos (r\phi)d\phi\nonumber\\
	&=&\frac{2\sin(r\phi_{+})}{\pi r}.
\end{eqnarray}
When $r=0$, the function is
\begin{eqnarray}
	G_{r}=\frac{1}{\pi}\int^{\phi_{+}}_0 d\phi-\frac{1}{\pi}\int^{\pi}_{\phi_{+}}d\phi=\frac{2\phi_{+}}{\pi}-1.
\end{eqnarray}
Combining Eqs. (19) and (20), we can obtain the formula of $G_r$ in the SL-I phase
\begin{eqnarray}
	G_{r}=\frac{2\sin(r\phi_{+})}{\pi r}-\delta_{0r}.
\end{eqnarray}

In the SL-II phase, there are two Fermi points $\phi_{+}$ and $\phi_{-}$, which divide the function $z(\phi)$ into three parts where $z(\phi)>0$ in the region $\phi\in[\phi_{+},\phi_{-}]$ and $z(\phi)<0$ in the regions $\phi\in[0,\phi_{+})\cup (\phi_{-},\pi]$. Then the kernel function in the SL-II phase has the form
\begin{eqnarray}
	F(\lambda,\alpha, r)&=&\frac{\cos (r\phi)[\lambda-\cos(\phi)-\alpha\cos(2\phi)]}{\sqrt{[\lambda-\cos(\phi)-\alpha\cos(2\phi)]^2}}\nonumber\\
	&=&\begin{cases}
		-\cos (r\phi),& \phi\in[0,\phi_{+})\\
		\cos (r\phi),& \phi\in[\phi_{+},\phi_{-}]\\
		-\cos (r\phi).& \phi\in(\phi_{-},\pi]
	\end{cases}
\end{eqnarray}
According to Eq. (13), we can obtain the $G_r$ function in the case of $r\neq 0$
\begin{eqnarray}
	G_{r}&=&\frac{1}{\pi}\int^{\phi_{+}}_0 \cos (r\phi)d\phi-\frac{1}{\pi}\int^{\phi_{-}}_{\phi_{+}} \cos (r\phi)d\phi\nonumber\\
	&+&\frac{1}{\pi}\int^{\pi}_{\phi_{-}} \cos (r\phi)d\phi\nonumber\\
	&=&\frac{2[\sin(r\phi_{+})-\sin(r\phi_{-})]}{\pi r},
\end{eqnarray}
and the formula in the case of $r=0$ is
\begin{eqnarray}
	G_{r}&=&\frac{1}{\pi}\int^{\phi_{+}}_0 d\phi-\frac{1}{\pi}\int^{\phi_{-}}_{\phi_{+}}d\phi+\frac{1}{\pi}\int^{\pi}_{\phi_{-}}d\phi\nonumber\\
	&=&\frac{2(\phi_+-\phi_-)}{\pi}+1.
\end{eqnarray}
Combining Eqs. (23) and (24), we can obtain the analytical formula of $G_r$ function in the SL-II phase
\begin{eqnarray}
	G_{r}=\frac{2[\sin(r\phi_{+})-\sin(r\phi_{-})]}{\pi r}+\delta_{0r}.
\end{eqnarray}

The expressions in Eqs. (17), (21) and (25) constitute the analytical form of the $G_r$ function in the XXT model, which is given in Eq. (7) of the main text. Based on the analytical formula, we can expediently obtain the two-site reduced state $\rho_r$ and study further the properties of long-range two-site quantum resources in the reduced states, such as two-site quantum coherence, quantum entanglement and quantum discord.

\subsection{IV. The confirmation of long-range two-site entanglement in the XXT model}

In multipartite spin systems, two-site entanglement is often short-range along with two-site distance and exhibits the phenomenon of entanglement sudden death \cite{ty2009sci}. Here, we study the entanglement property of the reduced state $\rho_r$ in the XXT model and analyze its long-range behavior of two-site entanglement. As quantified by the concurrence \cite{woo1998prl}, we have
\begin{equation}
C(\rho_r)={\rm{max}}\{0,2(|y^{+}|-\sqrt{u^{+}u^{-}})\},
\end{equation}
where $y^{\pm}$ and $u^{\pm}$ are the off-diagonal and diagonal elements of reduced state $\rho_r$ which can be calculated by the analytical formula of $G_r$. We mainly consider the spin-liquid phases in the XXT model, since the ground state in the ferromagnetic phase is product states for which the concurrence of reduced state is zero.

After some calculation, we can obtain the two-site concurrence in the SL-I phase
\begin{eqnarray}
	C(\rho_r)&=&{\rm{max}}\{0,2|y^{+}|-\frac{2}{\pi^2}\sqrt{(C_1-E_1)(D_1-E_1)}\}\nonumber\\
	&\approx&{\rm{max}}\{0,C_{l_1}(\rho_r)-\frac{2}{\pi^2}\sqrt{C_1\cdot D_1}\},
\end{eqnarray}
where the parameters are $C_1=\phi^2_{+}$, $D_1=(\pi-\phi_{+})^2$ and $E_1=\sin^2(r\phi_{+})/r^2$ with $\phi^+$ being the Fermi point, and, in the second equation, we use the two-site coherence $C_{l_1}(\rho_r)=2|y^+|$ in Eq. (6) of the main text and the approximation that the parameter $E_1$ tends to zero in the long-range case of two-site distance. Therefore, we know that the two-site concurrence is less than the corresponding two-site coherence. Furthermore, since the coherence is long-range and decays in the asymptotical mode, we can infer the result that the two-site entanglement in Eq. (27) can be long-range in the case of $C_1\cdot D_1$ tending to zero.

Next, we analyze the two-site entanglement in the SL-II phase and, according to the formula in Eq. (26), we can derive
\begin{eqnarray}
	C(\rho_r)&=&{\rm{max}}\{0,2|y^{+}|-\frac{2}{\pi^2}\sqrt{(E_2-G_2)(F_2-G_2)}\}\nonumber\\
	&\approx&{\rm{max}}\{0,C_{l_1}(\rho_r)-\frac{2}{\pi^2}\sqrt{E_2\cdot F_2}\},
\end{eqnarray}
where the parameters are $E_2=(\pi+\phi_{+}-\phi_{-})^2$, $F_2=(\phi_{+}-\phi_{-})^2$ and $G_2=[\sin(r\phi_{+})-\sin(r\phi_{-})]^2/r^2$ with $\phi^\pm$ being the two Fermi points, and we use the approximation that the parameter $G_2$ tends to zero in the long distance. After a similar analysis as that in the SL-I phase, we can obtain the result that the two-site entanglement is long-range in the case of
$E_2\cdot F_2$ tending to zero.

\begin{figure}
	\epsfig{figure=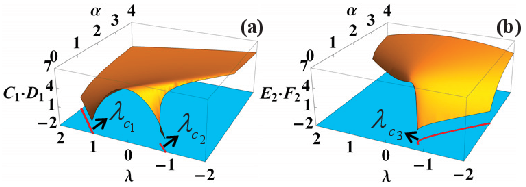,width=0.48\textwidth}
	\caption{(Color online) $C_1\cdot D_1$ (a) and $E_2\cdot F_2$ (b) as the functions of the parameters $\alpha$ and $\lambda$, where the red lines on the bottom indicate the zero values for the two functions and correspond to the critical lines $\lambda_1$, $\lambda_2 (\alpha\leq 0.25)$ and $\lambda_3$.}
\end{figure}

In Eqs. (27) and (28), the functions $C_1\cdot D_1$ and $E_2\cdot F_2$ are independent of the two-site distance. As shown in Fig. 1, we plot the two functions along with the change of the parameters $\alpha$ and $\lambda$, where the red lines in the bottom indicate the cases of $C_1\cdot D_1 = 0$ and $E_2\cdot F_2 = 0$, and the three red lines just correspond to the critical lines $\lambda_{c_1}=\alpha+1$, $\lambda_{c_2}=\alpha-1$ with the parameter $(\alpha\leq 0.25)$ and $\lambda_{c_3}={-(1+8\alpha^2)}/(8\alpha)$, respectively. It should be noted that the two-site coherence $C_{l_1}$ is zero when the functions $C_1\cdot D_1$ or $E_2\cdot F_2$ are zero. Therefore the long-range two-site entanglement occurs at the regions of the two functions tending to zero. In the main text, we choose three sets of parameters and confirm the existence of long-range two-site entanglement as shown in Fig. 3.

\subsection{V. Characterization of quantum phases via the long-range two-site quantum discord in the XXT model}

Quantum discord is a kind of typical quantum correlation in quantum information processing and can be written as \cite{ho2001prl,lh2001jpa}
\begin{eqnarray}
	 D_A(\rho_{AB})=S(\rho_{A})-S(\rho_{AB})+\underset{\{E^A_{k}\}}{\rm{min}}\underset{k}{\sum}p_{k}S(\rho_{B|k}),
\end{eqnarray}
where the measurement $\{E_k^A\}$ is performed on subsystem $A$ with the minimum running over all the projection measurements,  $S(\sigma)=-\mbox{Tr}\sigma\mbox{log}\sigma$ is the von Neumann entropy and $\rho_{B|k}=\mbox{Tr}_A[E_k^A\rho_{AB}E_k^A/\mbox{Tr}(E_k^A\rho_{AB}E_k^A)]$ is the output state of subsystem $B$ after the measurement $E_k^A$ with the probability being $p_k=\mbox{Tr}(E_k^A\rho_{AB}E_k^A)$.

In the XXT model, the two-qubit reduced state $\rho_{AB}=\rho_r$ has the form given in Eq. (2) of the main text, for which the single qubit reduced state for subsystem $A$ is
\begin{equation}
\rho_A=\left(
         \begin{array}{cc}
           u^{+}+z & 0 \\
           0 & u^{-}+z \\
         \end{array}
       \right).
\end{equation}
After some derivation, we can obtain the von Neumann entropies for the reduced states $\rho_A$ and $\rho_{AB}$ which can be written as
\begin{eqnarray}
	S(\rho_{A})&=&-Tr\rho_A\log_2\rho_A\nonumber\\
	&=&-\frac{1}{2}(1+\langle\sigma^z\rangle)\log_2[\frac{1}{2}(1+\langle\sigma^z\rangle)]\nonumber\\
	&&-\frac{1}{2}(1-\langle\sigma^z\rangle)\log_2[\frac{1}{2}(1-\langle\sigma^z\rangle)],
\end{eqnarray}
and
\begin{eqnarray}
	S(\rho_{AB})&=&-Tr\rho_{r}\log_2\rho_{r}\nonumber\\
	&=&-z_1\log_2z_1-z_2\log_2z_2\nonumber\\
	&&-u^{+}\log_2u^{+}-u^{-}\log_2u^{-},
\end{eqnarray}
where $z_1=z+y^+=(1+2\langle\sigma^x_0\sigma^x_r\rangle-\langle\sigma^z_0\sigma^z_r\rangle)/4$, $z_2=z-y^+=(1-2\langle\sigma^x_0\sigma^x_r\rangle-\langle\sigma^z_0\sigma^z_r\rangle)/4$ and $u^{\pm}=(1\pm 2\langle\sigma^z\rangle+\langle\sigma^z_0\sigma^z_r\rangle)/4$, respectively.

\begin{figure}
	\epsfig{figure=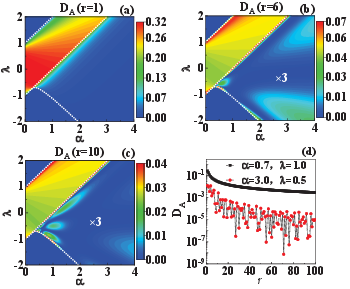,width=0.5\textwidth}
	\caption{(Color online) The change of quantum discord $D_A(\alpha,\lambda)$ for different two-site distances: (a) $r=1$, (b) $r=6$, and (c) $r=10$, respectively. In panel (d), two sets of parameters are chosen in SL-I and SL-II phases, where the asymptotical and oscillating decay modes of $D_A$ can serve as the diagnostic of quantum phases.}
\end{figure}

The third term in Eq. (29) is the measurement-induced conditional entropy, where the optimal measurement on subsystem $A$ can be determined via the method presented by Chen \emph{et al} \cite{qc11pra}. The two-qubit reduced state $\rho_{AB}=\rho_r$ in the XXT model has the $X$-shape for which the optimal measurement for $D_A(\rho_{AB})$ is $\sigma_z$ if
\begin{equation}
	\begin{split}
		(y^{+})^2\leqslant(u^{+}-z)(u^{-}-z),
	\end{split}
\end{equation}
and the optical measurement is $\sigma_x$ if
\begin{equation}
	\begin{split}
		|\sqrt{u^{+}u^{-}}-\sqrt{z^2}|\leqslant|y^{+}|,
	\end{split}
\end{equation}
where the parameters $y^\pm$, $u^\pm$ and $z$ are the matrix elements of $\rho_r$ given in Eq. (2) of the main text. Therefore, the measurement-induced conditional entropy in Eq. (29) is
\begin{equation}
	 \widetilde{S}(B|A)=\underset{\{\sigma_x^k~\texttt{or}~\sigma_z^k\}}{\rm{min}}\underset{k}{\sum}p_{k}S(\rho_{B|k}),
\end{equation}
where $\sigma_x^k$ and $\sigma_z^k$ with $k=0$ and $1$ are corresponding measurement operators for the Pauli measurements, and the related conditional entropies are
\begin{equation}
	\begin{split}
		 \underset{\{\sigma^k_x\}}{\sum}p_{k}S(\rho_{B|k})=-\lambda_+\log_2(\lambda_+)-\lambda_-\log_2(\lambda_-),
	\end{split}
\end{equation}
in which the parameters are $\lambda_{\pm}=(1\pm\sqrt{\langle\sigma^x_{0}\sigma^x_{r}\rangle^2+\langle\sigma^z\rangle^2})/2$, and
\begin{equation}
	\begin{split}
		 \underset{\{\sigma^k_z\}}{\sum}p_{k}S(\rho_{B|k})=-\sum_{i=+,-}p_{i}(\xi_i\mbox{log}_2\xi_i+\eta_i\mbox{log}_2\eta_i),
	\end{split}
\end{equation}
in which the probabilities and the corresponding eigenvalues of subsystem $B$ are
\begin{eqnarray}
	p_\pm&=&(1\pm\langle\sigma^z\rangle)/2,\nonumber\\
	\xi_\pm&=&(1-\langle\sigma^z_{0}\sigma^z_{r}\rangle)/[2(1\pm\langle\sigma^z\rangle)],\nonumber\\
	 \eta_\pm&=&(1\pm2\langle\sigma^z\rangle+\langle\sigma^z_{0}\sigma^z_{r}\rangle)/[2(1\pm\langle\sigma^z\rangle)].
\end{eqnarray}
Combining Eq. (35) and Eqs. (31)-(32), we can calculate the two-site quantum discord for the reduced state $\rho_r$ in the XXT model.

Next, we study the long-range property of quantum discord in the XXT model. As shown in Fig. 2, the change of $D_A(\alpha,\lambda)$ along the parameters $\alpha$ and $\lambda$ for the spin-spin distance $r=1, 6$ and $10$ are plotted in panels (a)-(c), where the quantum discord in SL-II phase is magnified by a factor of 3 in panels (b) and (c). The quantum correlation $D_A$ is zero in the ferromagnetic phase, and the nonzero $D_A$ can distinguish SL-I and SL-II phases from the ferromagnetic phase. Moreover, similar to the case of two-site coherence in the main text, the change pattern of quantum discord is different in the two spin-liquid phases as shown in panels (a)-(c), where $D_A$ is asymptotical in SL-I phase and oscillating in SL-II phase. The intrinsic reason is that $D_A$ decays along with the distance $r$ in the asymptotical mode for the SL-I phase and in the oscillating mode for SL-II phase, which can serve as an effective diagnostic tool for the two phases. In panel (d), we choose the parameters $(\alpha=0.7, \lambda=1.0)$ in SL-I phase and $(\alpha=3.0, \lambda=0.5)$ in SL-II phase, where the quantum discord exhibits distinctly different decay modes.

\subsection{VI. Detection of the TQPTs by the $l_1$-norm measure of quantum coherence}

In the main text, we pointed out that the first-order derivative of adjacent two-site coherence $C_{l_1}(\rho_r)$ can detect the topological quantum phase transition in the extended Ising model. Here, we first consider the TQPTs driven by the three-spin interaction $\alpha$, and the other parameters in the model are chosen to be $\gamma=1$, $\delta=-1$ and $\lambda=1$, respectively. In Fig. 3(a) and 3(b), we plot the $l_1$-norm coherence and its first-order derivative as a function of the parameter $\alpha$ for the adjacent spin pair. As shown in the figures, the quantum coherence $C_{l_1}(\rho_r)$ is a continuous function, but its first-derivative exhibits the divergent behaviors where the divergent points just indicate the critical points $\alpha_{c_1}=(-\sqrt{5}-1)/2$, $\alpha_{c_2}=0$, $\alpha_{c_3}=(\sqrt{5}-1)/2$ and $\alpha_{c_4}=2$ for the TQPTs.

\begin{figure}
	\epsfig{figure=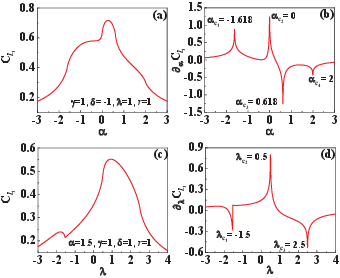,width=0.5\textwidth}
	\caption{(Color online) The $l_1$-norm of quantum coherence $C_{l_1}$ and its derivative $\partial_{\alpha}C_{l_1}$ versus the three-spin interaction $\alpha$ [panels (a) and (b)] and external magnetic field $\lambda$ [panels (c) and (d)] for adjacent two-spin pair.}
\end{figure}

\begin{figure*}
	\epsfig{figure=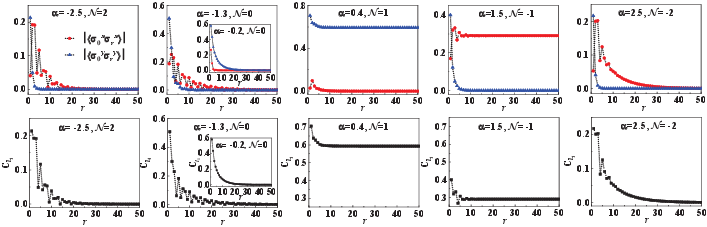,width=1.0\textwidth}
	\caption{(Color online) Spin correlations (the first row) and quantum coherence (the second row) as a function of two-site distance in topological phases with winding numbers $\mathcal{N}=2, 0, 1, -1$ and $-2$, where the TQPTs are driven by three-spin interaction $\alpha$ (other parameters are chosen to be $\gamma=1$, $\delta=-1$ and $\lambda=1$) and the long-range $l_1$-norm coherence is dominated by spin correlation $\langle \sigma_0^x\sigma_r^x\rangle$ (red-dot line) or $\langle \sigma_0^y\sigma_r^y\rangle$ (blue-triangle line) in different phases.}
\end{figure*}

We also analyze the TQPTs driven by external magnetic field $\lambda$, and the other parameters in the model are chosen to be $\alpha=1.5$, $\gamma=1$ and $\delta=1$, respectively. In Fig. 3(c) and 3(d), we plot the $l_1$-norm coherence and its first-order derivative as a function of the parameter $\lambda$ for the adjacent spin pair. As shown in the figures, the divergent points of the first-order derivative can indicate the critical points $\lambda_{c_1}=-1.5$, $\lambda_{c_2}=0.5$ and $\lambda_{c_3}=2.5$ for the TQPTs in the extended Ising model. Moreover, similar to the above cases, the TQPTs driven by the parameters $\gamma$ and $\delta$ can also be detected by the first-order derivative of $l_1$-norm quantum coherence for the adjacent two-spin pair.

\subsection{VII. The relation between spin correlations and quantum coherence in topological quantum phases}

According to Eq. (11) of the main text, the long-range two-site quantum coherence $C_{l_1}(\rho_r)$ in the extended Ising model can be expressed as the function of spin correlations $\langle \sigma_0^x\sigma_r^x\rangle$ and $\langle \sigma_0^y\sigma_r^y\rangle$. It is pointed out in the main text that the behaviors of long-range quantum coherence (such as the oscillating, asymptotical and freezing modes) can serve as the diagnostic tools for topological phases with different winding numbers. Here, we further study the relation between spin correlations and quantum coherence in the extended Ising model.

For the TQPTs driven by the three-spin interaction, we choose six typical values for the parameter $\alpha$, which correspond to the topological phases with the winding numbers $\mathcal{N}=0,\pm 1,\pm 2$, respectively. As shown in Fig. 4, two-site spin correlations  $\langle \sigma_0^x\sigma_r^x\rangle$ (red-dot lines) and $\langle \sigma_0^y\sigma_r^y\rangle$ (blue triangle lines) in five topological phases are plotted along with the increasing spin-spin distance in the first row of the figure, and the corresponding $l_1$-norm coherence $C_{l_1}(\rho_r)$s (black-square lines) in different phases are plotted in the second row of the figure. In the topological phase  $\mathcal{N}=2$, the quantum coherence decays in the oscillating mode (see the first panel in the second row) and mainly comes from the spin correlation $|\langle \sigma_0^x\sigma_r^x\rangle|$, where the spin correlation $|\langle \sigma_0^y\sigma_r^y\rangle|$ will disappear after a short two-site distance (see the first panel in the first row). The coherence in the topological phase $\mathcal{N}=0$ can decay in both oscillating and asymptotical modes, where the oscillating-damped coherence comes from spin correlation $|\langle \sigma_0^x\sigma_r^x\rangle|$, but the asymptotical-damped coherence is due to the spin correlation $|\langle \sigma_0^y\sigma_r^y\rangle|$ (see the panels and its insets in the second column of Fig. 4). In the topological phase $\mathcal{N}=-2$, the quantum coherence decays in the asymptotical mode and is mainly contributed by the spin correlation $\langle \sigma_0^x\sigma_r^x\rangle$, as shown in the two panels in the fifth column of Fig. 4.

Moreover, it was pointed out in the main text that the quantum coherence in topological phases $\mathcal{N}=\pm 1$ will exhibit freezing phenomenon after a short-range distance. In the panels of the third and forth columns of Fig. 4, we plot the quantum coherence and its corresponding spin correlations, where we find that the freezing coherence in the case of $\mathcal{N}=1$ is contributed by the spin correlation $|\langle \sigma_0^y\sigma_r^y\rangle|$ (red-dot line) but the freezing resource in the case of $\mathcal{N}=-1$ comes from the spin correlation $|\langle \sigma_0^x\sigma_r^x\rangle|$ (blue-triangle line).

\subsection{VIII. Behaviors of long-range quantum discord in the topological phases driven by three-spin interactions}

In the main text, we have analyzed the properties of different topological phases with the behaviors of long-range two-site quantum coherence. Here, we will further demonstrate that the behaviors of long-range quantum discord have the same functionality for characterization on the topological phases driven by three-spin interactions. The expression of two-site reduced state $(\rho_r=\rho_{AB})$ is given in Eq. (2) of main text, and the related quantum discord $D_A$ can be calculated according to the formula in Eq. (14) of main text where the von Neumann entropies $S(\rho_A)$ and $S(\rho_{AB})$ can be obtained via some direct calculation but the measurement-induced conditional entropy
\begin{equation}
\widetilde{S}(B|A)=\underset{\{E^A_{k}\}}{\rm{min}}\underset{k}{\sum}p_{k}S(\rho_{B|k})
\end{equation}
needs to be optimized by all the projective measurements $\{E_k^A\}$ on the subsystem $A$.

Although the two-site reduced state $\rho_{r}$ is still $X$-shape in the extended Ising model ($\gamma\neq 0$ and $\delta\neq 0$), the method presented in Ref. \cite{qc11pra} for selecting the optimal measurement of $D_A(\rho_{AB})$ is non-applicable due to the condition on matrix elements $\left|y^+ +y^-\right| \geq \left|y^+ -y^-\right|$ not being satisfied. Therefore, in order to calculate the measurement-induced conditional entropy in quantum discord, we will consider the situation for all the projective measurements in which the measurement operator can be written as \cite{mss09pra,ycl2011pra}
\begin{equation}
	\begin{split}
		E^{A}_i=V\ket{i}\bra{i}V^{\dag},
	\end{split}
\end{equation}
where $\{\ket{i}\}$ is the standard computational basis $\{\ket{0},\ket{1}\}$ and the unitary transformation matrix $V$ has the form
\begin{equation}
	\begin{split}
		&V=\begin{pmatrix} \cos\frac{\theta}{2} & e^{-i\phi}\sin\frac{\theta}{2} \\
			e^{i\phi}\sin\frac{\theta}{2} & -\cos\frac{\theta}{2}  \end{pmatrix},
	\end{split}
\end{equation}
in which the parameters satisfy $0\leq\theta\leq\pi$ and $0\leq\phi<2\pi$, respectively. In this case, the quantum discord $D_A$ can be obtained via the optimization of parameters  $\theta$ and $\phi$ in measurement-induced conditional entropy. After carefully numerical verifications similar to that of Ref. \cite{ycl2011pra}, we find that the optimal projective measurements in Eq. (39) correspond to the following three cases
\begin{eqnarray}
&&E_{k(\texttt{I})}^A:~~\theta=\pi/2,~\phi=\pi/2,\nonumber\\
&&E_{k(\texttt{II})}^A:~~\theta=\pi/2,~\phi=0,\nonumber\\
&&E_{k(\texttt{III})}^A:~~\theta=0,~\phi=0,
\end{eqnarray}
with $k=1,2$, and then the measurement-induced conditional entropy is
\begin{equation}
	\begin{split}
\widetilde{S}(B|A)=
\underset{\{E_{k(\texttt{I})}^A,E_{k(\texttt{II})}^A,E_{k(\texttt{III})}^A\}}{\rm{min}}\underset{k}{\sum}p_{k}S(\rho_{B|k}).
	\end{split}
\end{equation}
In Fig. 5, we choose three typical quantum states of $\rho_r$ for which the optimal measurements for $\widetilde{S}(B|A)$ correspond to
$E_{k(\texttt{I})}^A, E_{k(\texttt{II})}^A$ and $E_{k(\texttt{III})}^A$, respectively.

\begin{figure}
	\epsfig{figure=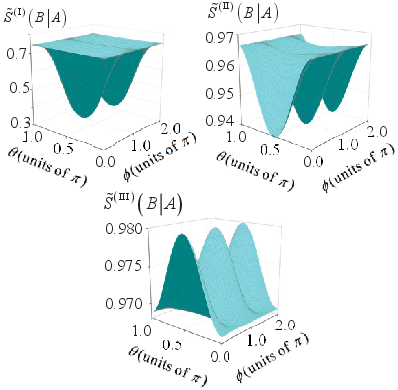,width=0.48\textwidth}
	\caption{(Color online) The measurement-induced conditional entropies $\widetilde{S}(B|A)$ for three typical quantum states $\rho_r$ ($r=2$) which correspond to optimal projective measurements: (a) $\{E_{k(\texttt{I})}^A\}$ for $\rho_r$ with parameters $\gamma=1$, $\delta=-1$, $\lambda=1$ and $\alpha=0.1$, (b) $\{E_{k(\texttt{II})}^A\}$ for $\rho_r$ with parameters $\gamma=1$, $\delta=-1$, $\lambda=1$ and $\alpha=-2.5$, and (c) $\{E_{k(\texttt{III})}^A\}$ for $\rho_r$ with parameters $\gamma=1$, $\delta=-2$, $\lambda=-0.3$ and $\alpha=1$, respectively.}
\end{figure}

Based on the previous analysis, we can derive the two-site quantum discord
\begin{equation}
D_A(\rho_r)=\widetilde{S}(B|A)-S(B|A),
\end{equation}
where the quantum condition entropy $S(B|A)=S(AB)-S(A)$ and the measurement-induced conditional entropy $\widetilde{S}(B|A)$ is given by Eq. (43) in which
\begin{eqnarray}
\widetilde{S}^{(\texttt{I})}(B|A)&=&\underset{E_{k(\texttt{I})}^A}{\sum}p_{k}S(\rho_{B|k})\nonumber\\
&=&-\beta_+\log_2(\beta_+)-\beta_-\log_2(\beta_-)
\end{eqnarray}
with $\beta_{\pm}=[1\pm(\langle\sigma^y_{0}\sigma^y_{r}\rangle^2+\langle\sigma^z\rangle^2)^{1/2}]/2$,
\begin{eqnarray}
\widetilde{S}^{(\texttt{II})}(B|A)&=&\underset{E_{k(\texttt{II})}^A}{\sum}p_{k}S(\rho_{B|k})\nonumber\\
&=&-\lambda_+\log_2(\lambda_+)-\lambda_-\log_2(\lambda_-)
\end{eqnarray}
with $\lambda_{\pm}=[1\pm(\langle\sigma^x_{0}\sigma^x_{r}\rangle^2+\langle\sigma^z\rangle^2)^{1/2}]/2$, and
\begin{eqnarray}
\widetilde{S}^{(\texttt{III})}(B|A)&=&\underset{E_{k(\texttt{III})}^A}{\sum}p_{k}S(\rho_{B|k})\nonumber\\
&=&-\sum_{i=+,-}p_{i}(\xi_i\mbox{log}_2\xi_i+\eta_i\mbox{log}_2\eta_i)
\end{eqnarray}
with $p_\pm=(1\pm\langle\sigma^z\rangle)/2$, $\xi_\pm=(1-\langle\sigma^z_{0}\sigma^z_{r}\rangle)/[2(1\pm\langle\sigma^z\rangle)]$ and $\eta_\pm=(1\pm2\langle\sigma^z\rangle+\langle\sigma^z_{0}\sigma^z_{r}\rangle)/[2(1\pm\langle\sigma^z\rangle)]$, respectively.

In Fig. 6(a)-(c), we plot quantum discord $D_A(\rho_r)$ as a function of three-spin interaction $\alpha$ and two-site distance $r$, where the different topological phases are labeled by the corresponding winding numbers. In the case of $\mathcal{N}=2$, the quantum discord decays in the oscillating mode along with two-site distance for a given value of $\alpha$. However, in the case of $\mathcal{N}=-2$, quantum discord decays in the asymptotical mode after a short two-site distance as shown in Fig. 6(c). Moreover, the quantum discord in topological phase $\mathcal{N}=0$ can decay in both oscillating and asymptotical modes as shown in Fig. 6(a). In particular, given a value of $\alpha$, the nonzero quantum discord in topological phases $\mathcal{N}=\pm1$ exhibits quantum correlation freezing phenomenon after a short two-site distance as shown in Fig. 6(b). In generic case, the freezing quantum discord in topological phase $\mathcal{N}=1$ has a larger value than that of quantum discord in topological phase $\mathcal{N}=-1$. In Fig. 6(d)-(f), we select some typical values in different topological phases and plot the long-range behaviors of quantum discord, where the oscillating-damped, asymptotical-damped and freezing modes of $D_A(\rho_r)$ can serve as the effective diagnostic tools for quantum phases. It should be noted that the steady quantum discord is a kind of topologically protected long-range quantum correlation and has the potential applications in quantum computation and quantum communication.

\begin{figure}
	\epsfig{figure=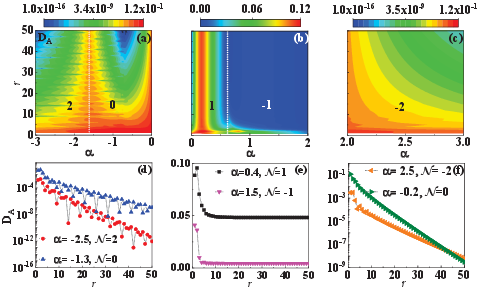,width=0.5\textwidth}
	\caption{(Color online) (a)-(c) The quantum discord $D_A(\rho_r)$ as a function of three-spin interaction $\alpha$ and two-site distance $r$ in different topological phases with winding numbers $\mathcal{N}=0,\pm1$ and $\pm2$ where the other parameters are chosen to be $\gamma=1$, $\delta=-1$, and $\lambda=1$, respectively. (d)-(f) The behaviors of long-range two-site discord for some typical values of $\alpha$ in the five quantum phases.}
\end{figure}

\begin{figure*}
	\epsfig{figure=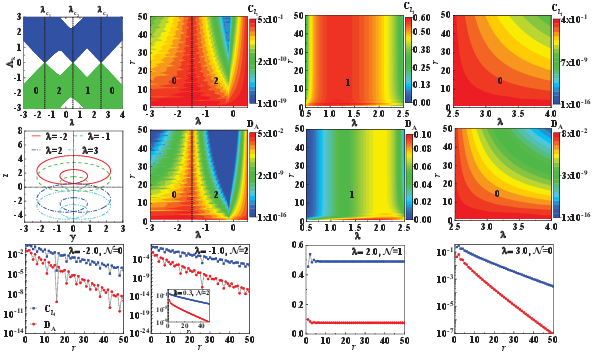,width=1.0\textwidth}
	\caption{(Color online) Long-range properties of quantum coherence (the first row) and quantum discord (the second row) as a function of two-site distance and external magnetic field accompanied with the corresponding energy spectra $\Lambda_k(\lambda)$ and trajectories of winding vectors, and, in the third row, the oscillating-damped, asymptotical-damped and freezing two-site quantum resources can serve as a set of effective diagnostic tools for topological phases in the extended Ising model.}
\end{figure*}

\subsection{IX. Behaviors of long-range two-site quantum resources in the topological phases driven by external magnetic field}

In this section, we will study the long-range behaviors of quantum coherence and quantum discord in the topological phases driven by external magnetic field $\lambda$, where the other parameters in the extended Ising model are chosen to be $\alpha=1.5, \gamma=1, \delta=1$, respectively. The critical points of the topological quantum phase transitions can be obtained via solving the characteristic equation \cite{spl2018pra}
\begin{equation}
3\xi^{-2}/2+\xi^{-1}-\lambda=0
\end{equation}
for which the solutions are $\lambda_{c_1}=-1.5$ corresponding to $\xi_1=\exp[\pm i\arccos(-1/3)]$, $\lambda_{c_2}=0.5$ corresponding to $\xi_2=-1$, and $\lambda_{c_3}=2.5$ corresponding to $\xi_3=1$, respectively \cite{ykw19qip}. The two-site quantum coherence can be calculated according to the formula in Eq. (11) of the main text, and the corresponding quantum discord can be obtained by the method given in Eq. (44) of the supplemental materials.

In Fig. 7, we plot the energy spectra of the system with a chain length $L=1001$ (the three critical points and four regions of topological phases are labeled), the trajectories of the winding number vectors in the $x$-$y$ planes (corresponding to winding numbers $\mathcal{N}=0,1$, and $2$), and the long-range properties of two-site quantum coherence and quantum discord. As shown in the first row of the figure, the $l_1$-norm coherence $C_{l_1}(\rho_r)$ in different regions of topological phases exhibit the oscillating-damped, asymptotical-damped and freezing modes along with two-site distance. In particular, the topologically protected quantum coherence in the phase $\mathcal{N}=1$ will attain to a steady value after a short two-site distance. The case for two-site quantum discord $D_A(\rho_r)$ is similar, which exhibits the oscillating-damped, asymptotical-damped and freezing behaviors along with two-site distance in different regions of topological phases as shown in the second row of the figure. In the third row of the figure, we choose some typical values of external magnetic field $\lambda$ and plot the long-range behaviors of quantum coherence and quantum discord, where the two kinds of quantum resources exhibit the similar decay modes which further demonstrate that these long-range behaviors of quantum resources can serve as a set of effective diagnostic tools for topological quantum phases. Moreover, similar to the case in the topological quantum phases $\mathcal{N}=\pm1$ driven by three-spin interaction $\alpha$, the topologically protected quantum resources in topological phase $\mathcal{N}=1$ driven by magnetic field $\lambda$ still exhibit the freezing phenomenon along with two-site distance.

\end{document}